\documentclass[aps,prb,amsmat,amssymb,superscriptaddress, twocolumn, showpacs]{revtex4}
\usepackage[english]{babel}
\usepackage{graphicx}
\usepackage{transparent}
\usepackage{amsmath}
\usepackage{amssymb}
\usepackage{epstopdf}
\usepackage{amsfonts}
\usepackage{bm}
\usepackage{times}
\usepackage{pdfpages}
\usepackage{mathtools}
\usepackage{hyperref}
\usepackage{tabularx}
\usepackage{hhline}
\usepackage[normalem]{ulem}

\newcommand{\+}{\dagger}
\newcommand{\Vab}{V}
\newcommand{\Vaa}{V}

\newcommand{\I}{\mathrm{i}}

\newcommand{\ave}[1]{\langle #1 \rangle}

\newcommand{\green}[2]{\langle \mathcal{T_C} #1(t) #2(t') \rangle}
\newcommand{\beq}[1]{\begin{equation} #1 \end{equation}}
\newcommand{\bsplit}[1]{\begin{equation} \begin{split} #1 \end{split} \end{equation}}

\newcommand{\astcycl}{\mathrlap{\kern0.085em{\circlearrowright}}\ast}
\newcommand{\taucycl}{\mathrlap{\kern0.42em{\bullet}}\circlearrowright}
\DeclareMathOperator{\Tr}{Tr}
\def\<{\left\langle}
\def\>{\right\rangle}
\newcommand{\vk}{\textbf{k}} 
\newcommand{\vq}{\textbf{q}}

\begin{document}

\title{Photo-induced Dirac cone flattening in BaNiS$_2$}

\author{Nikolaj Bittner}
\email{nikolaj.bittner@unifr.ch}
\affiliation{Department of Physics, University of Fribourg, 1700 Fribourg, Switzerland}
\author{Denis Gole\v{z}}
\affiliation{Jo\v zef Stefan Institute, Jamova 39, SI-1000, Ljubljana, Slovenia}
\affiliation{Faculty of Mathematics and Physics, University of Ljubljana, Jadranska 19, 1000 Ljubljana, Slovenia}
\author{Michele Casula}
\affiliation{Institut de Min\'eralogie, de Physique des Mat\'eriaux et de Cosmochimie (IMPMC), Sorbonne Universit\'e, CNRS UMR 7590, IRD UMR 206, MNHN, 4 place Jussieu, 75252 Paris, France}
\author{Philipp Werner}
\email{philipp.werner@unifr.ch}
\affiliation{Department of Physics, University of Fribourg, 1700 Fribourg, Switzerland}

\begin{abstract}
Using a real-time implementation of the self-consistent $GW$ method, we theoretically investigate the photo-induced changes in the electronic structure of the quasi two-dimensional semi-metal BaNiS$_2$. This material features four Dirac cones in the unit cell and our simulation of the time- and momentum-resolved nonequilibrium spectral function reveals a flattening of the Dirac bands after a photo-doping pulse with a 1.5 eV laser. The simulation results are consistent with the recently reported experimental data on photo-doped BaNiS$_2$ and ZrSiSe, another Dirac semi-metal. A detailed analysis of the numerical data allows us to attribute the nonequilibrium modifications of the Dirac bands to (i) an increased effective temperature after the photo-excitation, which affects the screening properties of the system, and (ii) to  
nontrivial band shifts in the photo-doped state, which are mainly induced by the Fock term. 
\end{abstract}

\pacs{71.10.Fd, 05.70.Ln}

\maketitle

\section{Introduction}

Photo-induced changes in band dispersions and quasi-particle life-times have been reported in different classes of solids. In charge-transfer insulators such as cuprates, for example, the photo-doping across the charge transfer gap results in substantial band shifts and band broadenings,~\cite{novelli2014,rameau2014,matsuda1994,okamoto2011,okamoto2010,cilento2018} which can be explained by the induced changes in the electrostatic energy (Hartree shifts)~\cite{cilento2018,sandri2015} and enhanced nonlocal charge fluctuations.\cite{Golez2019,golez2019a,tancogne2018} In correlated semiconductors, photo-excitation typically leads to a narrowing of the gap, which is referred to as band gap renormalization.\cite{wegkamp2014,pagliara2011photoinduced} However, recent experiments on excitonic systems revealed a richer behavior with a transient enhancement~\cite{Mor2017} or suppression~\cite{okazaki2018photo,tang2020,baldini2020} of the band gap, which was attributed to the underlying  excitonic order.\cite{murakami2017,tanaka2018,tanabe2018,ryo2019} 

\begin{figure}[b]
\centering
		  \includegraphics[
		  width=\columnwidth]{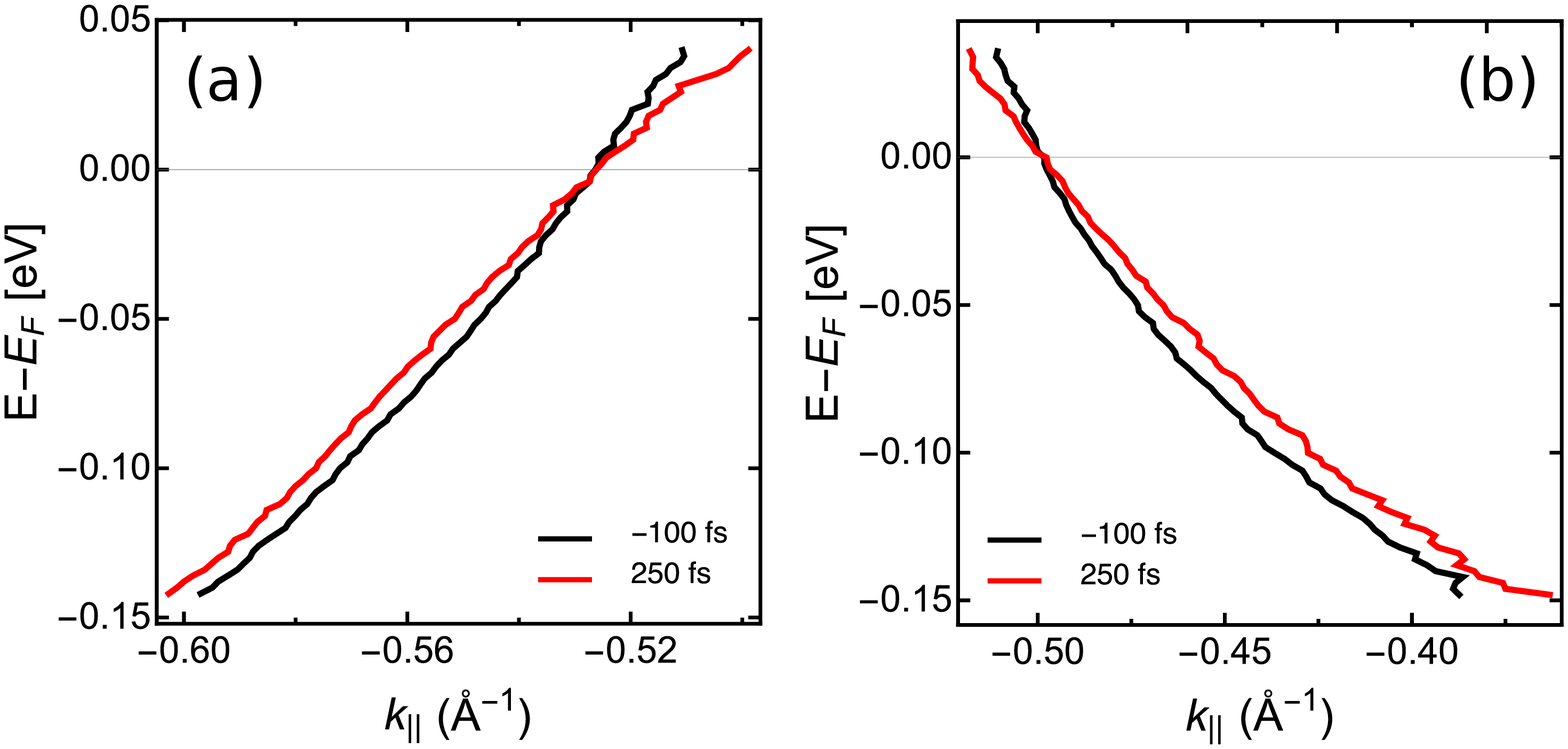}
\caption{Dirac cone flattening for the left~(a) and the right~(b) 
branch after photo-excitation with a 1.5 eV laser, measured with tr-ARPES. 
  The black lines show the equilibrium dispersion, and the red lines the dispersion measured after the photo-doping pulse.
The {\bf k}-points are aligned along the $\Gamma$-M direction, $\Gamma$ (M) being located at $k_\parallel=0$ ($k_\parallel=\pm 1.00$ \mbox{\AA}$^{-1}$) (Experimental data adapted from Ref.~\onlinecite{nilforoushan2019a}.)
}
\label{fig:Exp}
\end{figure}

In weakly or moderately correlated metals, electron-electron interactions typically result in a narrowing of the bands near the Fermi level, so that photo-excitation and its associated heating effects are naively expected to lead to a band widening. However, in the Dirac semi-metal BaNiS$_2$, a recent photo-doping study \cite{nilforoushan2019a} has observed  
a flattening of the Dirac cone in the photo-doped state (Fig.~\ref{fig:Exp}). Remarkably, this flattening of the Dirac cone persists up to 1 ps, indicating a bottleneck in the electronic dynamics. Similar measurements have also been reported in photo-doping experiments on ZrSiSe, a different material with Dirac dispersions.\cite{Gatti2020} These experimental results suggest a nontrivial effect of the photo-induced changes on the screening environment in these materials. Indeed, temperature-dependent equilibrium calculations\cite{nilforoushan2019a,Gatti2020} point to an important role of the nonlocal interactions in this unusual response of the electronic structure to photo-carriers. As an extreme case of photo-manipulation of relativistic materials, recent photo-emission studies have demonstrated an ultrafast Lifshitz transition in the correlated type-II Weyl semi-metal $T_d$-MoTe$_2$.\cite{beaulieu2021ultrafast}

To get deeper insights, a proper modeling and simulation of the nonequilibrium states in Dirac semi-metals is needed. Because these materials are moderately correlated, and the time-dependent screening of long-ranged interactions apparently plays an important role,  a promising approach is the nonequilibrium implementation of the $GW$ method.\cite{Hedin1965} This formalism has been successfully used to study transport in mesoscopic systems,\cite{thygesen2007} relaxation dynamics in various electron-boson coupled systems,\cite{sentef2013,rameau2016} as well as the nonthermal destruction of ordered states.\cite{golez2016} In combination with dynamical mean field theory,\cite{aoki2014} it allowed to reveal the effects of nonlocal charge fluctuations in photo-doped charge transfer insulators.\cite{Golez2019} However, nonequilibrium $GW$ studies of bulk materials have to the best of our knowledge been performed so far mainly in simple model contexts. 
Here, we bridge the gap to realistic nonequilibrium materials simulations by treating an {\it ab initio} 
model \cite{nilforoushan2019a, nilforoushan2019b} for BaNiS$_2$ within the framework of nonequilibrium $GW$, to study the effect of photo-excitation with a 1.5~eV pump pulse on the electronic structure and in particular the Dirac cone dispersion.

Our simulations reproduce the nontrivial band flattening, in 
qualitative agreement with the experimental findings based on time-resolved angle-resolved photoemission spectroscopy (tr-ARPES) (see Fig.~\ref{fig:Exp}). These encouraging results, in combination with the controlled many-body framework, allow us to address the central question of the paper: which many-body processes are mainly responsible for these photo-induced modifications in the electronic structure and which aspects are specific to Dirac systems? The direct access to the time-dependent self-energy and the screened interaction (or polarization) enables an analysis of the physical processes driving the photo-induced band renormalization of the Dirac states. In particular, we will clarify the effect of the photo-induced heating on the screening properties, and reveal the importance of the photo-induced Fock exchange modifications, which result from the nonthermally populated bands.

The paper is organized as follows. Sections~\ref{sec:Mod} and~\ref{sec:Med} introduce the realistic two-band model for BaNiS$_2$ and the real-time $GW$ method used in the numerical simulations, respectively. Section~\ref{sec:EqResults} discusses equilibrium properties of this correlated material, while Section~\ref{sec:NonEqResults} analyzes the time-dependent changes in the Dirac dispersion of BaNiS$_2$ after a photo-excitation. Section~\ref{sec:Summary} summarizes our findings. 

\section{Model}
\label{sec:Mod}
We consider a two-dimensional ($2d$) two-orbital system representing the low-energy electronic structure of BaNiS$_2$ at half-filling, as determined in Refs.~\onlinecite{nilforoushan2019a, nilforoushan2019b}.
Within this {\it ab-initio} derived description, we will focus on the charge dynamics, which  is relevant to study the renormalization of the Dirac states seen in tr-ARPES. Thus, we restrict
the model to spin-less fermions, and treat  BaNiS$_2$ 
by the Hamiltonian 
\beq{
\label{eq:H}
H(t)=H_\text{kin}+H_\text{int}+H_\text{dip}(t),
}
with a kinetic term of the tight-binding form
\bsplit{
\label{eq:Hkin}
	H_{\text{kin}}=&\sum_{\alpha, \vk} \Big[E_\alpha -\mu -t_\alpha (\cos(k_x)+\cos(k_y)) \\
	&\hspace{8mm}+ t'_\alpha \cos(k_x) \cos(k_y)\Big] n_{\vk \alpha}\phantom{\sum_{\alpha, \vk}} \\
	&+\sum_{\alpha\neq\beta,\vk} [ t_{\alpha,\beta} \sin(k_x)  \sin(k_y) d^{\dagger}_{\vk \alpha} d_{\vk \beta} + \text{H.c.} ]\,. 
}
Here, we have introduced the creation operators $d^{\dagger}_{\vk \alpha}$ for spin-less fermions 
in the orbitals $\alpha=z^2,x^2-y^2$ with momentum $\vk$, and $n_{\vk \alpha}=d^{\dagger}_{\vk a} d_{\vk a}$. The momentum is defined in a rotated Brillouin zone with basis vectors $\vk_1=(1,1)/\sqrt{2}$ and $\vk_2=(1,-1)/\sqrt{2}$. We will neglect the $z$ dependence of the dispersion. The hopping parameters and local energies in eV units are taken from Ref.~\onlinecite{nilforoushan2019b}:
\begin{equation}
\label{eq:parm}
	\begin{array}{lll}
E_{z^2} = -0.64, & \hspace{5mm}\mbox{}& E_{x^2-y^2} = 0.24,\\ 
t_{z^2} = -0.68, & & t_{x^2-y^2} = 0.28,\\
t^\prime_{z^2} = 0.66, & & t^\prime_{x^2-y^2} = -0.96,\\
t_{z^2,x^2-y^2} = 0.28, & & t_{x^2-y^2,z^2}=0.28.
	\end{array}
\end{equation}
The interaction term is restricted to density-density interactions of the form 
$H_{\text{int}}=\frac{1}{2}\sum_{i j}\sum_{\alpha \beta} \Vab_{|i-j|}n_{i\alpha} n_{j\beta} - \frac{1}{2}\sum_{i}\sum_{\alpha} \Vaa_\text{loc}n_{i\alpha} n_{i\alpha}$, which may be split into the non-local intraorbital interaction $\frac{1}{2}\sum_{i j}\sum_{\alpha \beta} \delta_{\alpha,\beta} \Vab_{|i-j|}n_{i\alpha} n_{j\beta} - \frac{1}{2}\sum_{i}\sum_{\alpha} \Vaa_\text{loc}n_{i\alpha} n_{i\alpha}$ and the interorbital interaction $\frac{1}{2}\sum_{i j}\sum_{\alpha \beta} (1-\delta_{\alpha,\beta}) \Vab_{|i-j|}n_{i\alpha} n_{j\beta}$. Hence, in momentum space, the interaction becomes
\bsplit{
\label{eq:Hint}
	H_{\text{int}}=&\frac{1}{2 N}
	\sum_{\vk,\vk',\vq} \sum_{\alpha \beta} \delta_{\alpha,\beta} \Vab(\vq) d_{\vk+\vq \alpha}^{\dagger} d_{\vk \alpha} d_{\vk'-\vq \beta}^{\dagger} d_{\vk' \beta} \\
	&- 
	\frac{1}{2 N}\sum_{\vk_1,\vk_2,\vk_3} \sum_{\alpha} \Vaa_\text{loc} d_{\vk_1+\vk_3-\vk_2\alpha}^{\dagger} d_{\vk_1 \alpha} d_{\vk_2 \alpha}^{\dagger} d_{\vk_3 \alpha}\\
	&+ 
	\frac{1}{2 N}\sum_{\vk,\vk',\vq} \sum_{\alpha \beta} (1-\delta_{\alpha,\beta}) \Vab(\vq) d_{\vk+\vq \alpha}^{\dagger} d_{\vk \alpha} d_{\vk'-\vq \beta}^{\dagger} d_{\vk' \beta},
}
where $N$ is the number of ${\bf k}$ points in the $2d$ Brillouin zone.

To obtain an interaction $\Vab$ which takes into account the $3d$ nature of the material, we will 
consider a system of stacked layers with interlayer distance $I_c$. In this setup, the interaction vertex of the 
layered $3d$ system is given by\cite{shung1986}
\beq{
	\Vab(\vq,k_z)= \frac{1}{4\pi\epsilon_0}\frac{2\pi e^2}{\kappa_0 |\vq|} \frac{\sinh(|\vq| I_c)}{\cosh(|\vq| I_c)-\cos(k_z I_c)}, \hspace{2mm}\forall \alpha, \beta.
\label{eq:Vqkz}
}
where $\vq=(q_x, q_y)$ denotes the momentum in the $2d$ layers, and $k_z$ the momentum perpendicular to the planes. 
For BaNiS$_2$, we take the in-plane lattice constant
$a=3.140$~\mbox{\AA} and $I_c=8.91$ \mbox{\AA}. 
The dimensionless momenta $q_xa$, $q_ya$ and $k_z I_c$ are in the range $[ -\pi, \pi]$.
$\kappa_0$ is the dielectric constant, which takes into account the screening from other bands, which have been discarded in the downfolding to the low-energy model.  
The constrained Random Phase Approximation (cRPA)\cite{aryasetiawan2012} interaction of the closely related compound BaCoS$_2$ suggests a screening of the bare interaction by $\kappa_0\approx 10$.\cite{Santos2018} Indeed, the interaction with $\kappa_0=10$ results in a flattening of the Dirac cones after the photo-excitation by $\approx 5\%$ (see Appendix~\ref{sec:kappa10}), which is in reasonable agreement with the experimental results (Fig.~\ref{fig:Exp}).\cite{nilforoushan2019a} 
However, for a better visualisation of the photo-induced effects, we choose in the following simulations 
$\kappa_0=6$, which results in stronger correlations and larger band shifts, without qualitatively changing the results.

\begin{figure}[t]
\centering
		  \includegraphics[
		  width=\columnwidth]{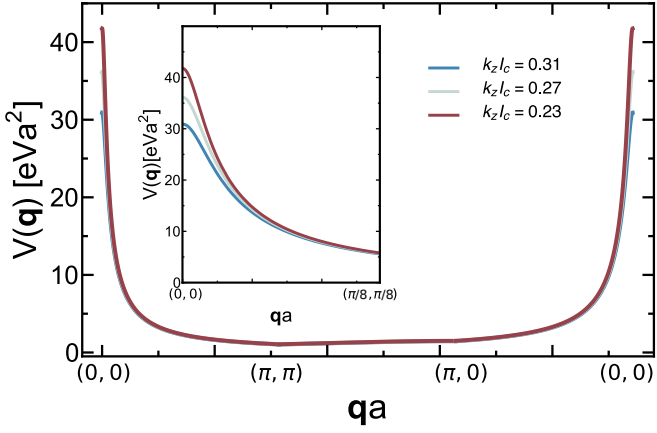}
\caption{Effective $2d$ interaction from Eq.~\eqref{eq:Veff} with $k_{z,\text{cut}}I_c=0.31$ (blue), $0.27$ (light blue), and 0.23 (dark red). Here, the momentum $\vq$ is scaled with the lattice constant $a$. 
}
\label{fig:Vkcut0012}
\end{figure}

From Eq.~(\ref{eq:Vqkz}), the effective interaction within a single $2d$ layer is obtained by averaging over $k_z$:
\bsplit{
\label{eq:Veff}
	\Vab(\vq)
	&=\frac{I_c}{\pi}\int^{\pi/I_c}_{k_{z,\text{cut}}} dk_z \Vab(\vq,k_z).
}
Because the Coulomb-like interaction $\Vab(\vq,k_z)$ has a singularity at $\vq=(0,0)$ and $k_z=0$, we introduce a small-momentum cutoff for the integration along the $k_z$ direction. This cuts off the long-ranged part of the interaction, $\Vab(\vq\rightarrow 0)$, as illustrated in Fig.~\ref{fig:Vkcut0012}. In the actual calculations, we use $k_{z,\text{cut}}I_c=0.23$. 
We have checked that this $k_{z,\text{cut}}$ is small enough that the results are not qualitatively affected. 

Due to the strong hybridization between the $d$ and $p$ orbitals in BaNiS$_2$, we describe the photo-excitation of the system by a laser-induced (dipole allowed) population transfer between the orbitals,
\beq{
	H_\text{dip}(t) = \sum_{\vk} A(t) d^\+_{\vk z^2} d_{\vk x^2-y^2} + \text{H.c.} \,.
\label{eq:Hdip}
}
Here, the pump pulse $A(t) = A_0 e^{-(t-t_0)^2/\tau^2}\sin(\omega(t-t_0))$ with frequency $\omega$ has a Gaussian envelope with a maximum $A_0$ at time $t_0$ and a full width at half maximum $\tau$. 

Energy and temperature will be measured in units of eV and time will be measured in units of 
fs.

\section{Method}
\label{sec:Med}

\subsection{Green functions}

We will treat the equilibrium and nonequilibrium properties of model (\ref{eq:H}) within the 
$GW$ approximation.\cite{golez2016} 
The time evolution of the correlated system is simulated using the {\tt NESSi} software library,\cite{schueler2020} which is based on the L-shaped Kadanoff-Baym contour formalism.\cite{aoki2014} 

It is convenient to introduce a spinor representation for the two orbitals:
\beq{
\Psi_{\vk}\equiv 
\begin{pmatrix}
d_{\vk,1}  \\
d_{\vk,2}  \\
\end{pmatrix}.
}
The corresponding 2$\times$2  Green's function is given by the contour $\cal{C}$ ordered expectation value:
\beq{
	\hat G_\vk(t,t')=-\I \green{\Psi_\vk}{\Psi_\vk^\+}
}
and is determined from the solution of a Dyson equation with appropriate self-energy. For later use, we also introduce the single-particle density matrix $\rho_{\vk,\alpha\beta}=\ave{\Psi^{\dagger}_{\vk,\beta} \Psi_{\vk,\alpha}}$ and its local component
$\rho_{\text{loc},\alpha\beta}=\frac{1}{N}\sum_\vk \rho_{\vk,\alpha\beta}$, where $N$ denotes the number of $\vk$ points in the $2d$ Brillouin zone. All the results shown will be calculated for $20\times 20$ $\vk$ points.

\subsection{Hartree-Fock approximation}
To calculate the Hartree and Fock self-energies, we employ the usual mean-field decoupling of the interaction term~\eqref{eq:Hint}. The 2$\times$2 Hartree self-energy in orbital space is then given by
\begin{align}
	\Sigma^H_{\vq,\alpha\beta}(t)=&\delta_{\alpha\beta}\Big( V(|\vq|=0) 
	 \rho_{\text{loc},\bar\alpha\bar\alpha}(t)\nonumber\\
	&\hspace{6mm}+ [\Vaa(|\vq|=0)	-\Vaa_\text{loc}] \rho_{\text{loc}, \alpha \alpha}(t)\Big),
\end{align}
where the first term corresponds to the inter-orbital interaction and the second term to the non-local intra-orbital interaction. An overline marks the opposite orbital.  

As is seen in Fig.~\ref{fig:Vkcut0012}, for small cutoff $k_{z,\text{cut}}$ the potential $\Vab(|\vq|=0)$ becomes very large. To avoid numerical problems resulting from this, we impose charge neutrality, so that the attractive potential of a neutralizing homogeneous background shifts the total Hartree term to zero.
In other words, $\Sigma^H_\vq \propto \< \rho_\mathrm{tot} \>$,
and $\< \rho_\mathrm{tot} \> = 0$, if the homogeneous background is taken into account. In practice, 
we set the Hartree self-energy in the initial equilibrium state to zero, while the orbitally-resolved components take nonzero values in the photo-excited nonequilibrium state.

The 2$\times$2 Fock self-energy term in orbital space is given by 
\bsplit{
\label{eq:Fock}
	\Sigma^F_{\vk,\alpha\bar\alpha}(t)&=-\frac{1}{N}
	\sum_\vq V(\vq)
	\rho_{\vk-\vq,\alpha\bar\alpha}(t),\\
	\Sigma^F_{\vk, \alpha\alpha}(t)&=-\frac{1}{N}\sum_\vq \Vaa(\vq) \rho_{\vk-\vq, \alpha\alpha}(t)+\Vaa_\text{loc}\rho_{\text{loc},\alpha\alpha}(t).
}

\begin{figure*}[t!]
\centering
		\includegraphics[width=0.95\textwidth]{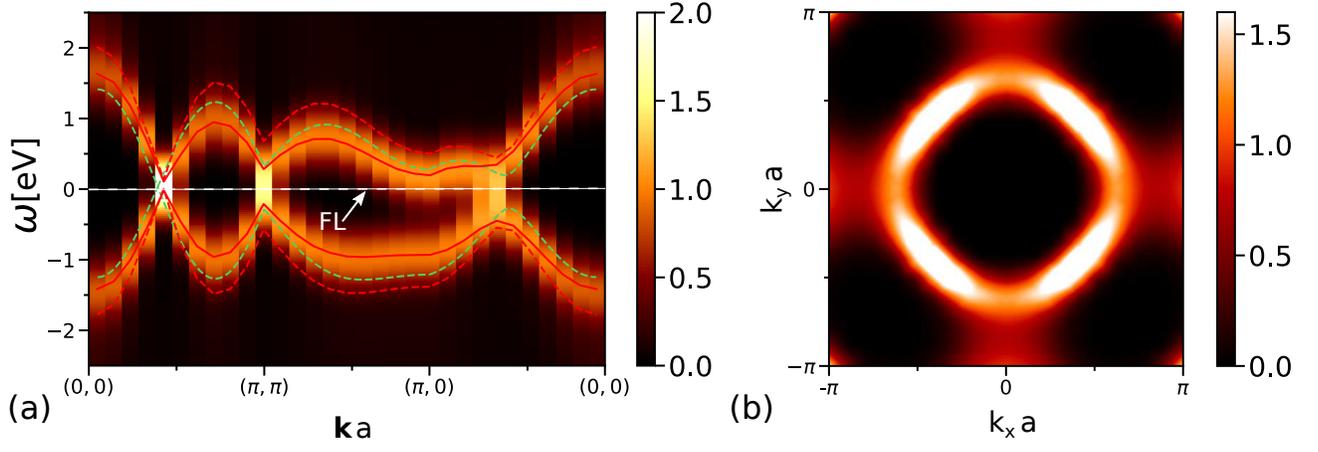}
\caption{(a) ${\bf k}$-dependent band dispersion $A_\vk(\omega)$ for BaNiS$_2$ calculated for $T=0.07\,$eV on a 20$\times$20 {\bf k} grid. For the Fourier transformation we use a time window $t_\mathrm{max}=6.8\,$fs. The intensity is measured in units $1/$eV.  The red dashed line represents the Hartree-Fock contributions to the band structure, while the solid red line shows the QPGW band structure defined by Eq.~(\ref{Eq:PeakPos}). For comparison, we also show the non-interacting band dispersion by the green dashed line. The horizontal dashed line corresponds to the Fermi level FL. (b) $A_\vk(\omega=0)$ plotted in the first Brillouin zone, showing the four Dirac points (surrounded by high intensity regions) on the diagonals. The intensity is measured in units $1/$eV. 
}
\label{fig:BandStruct}
\end{figure*} 

\subsection{GW approximation}

In the $GW$ method\cite{Hedin1965} the effect of screening is taken into account at the Random Phase Approximation level. This formalism can be derived from a Luttinger-Ward functional and hence yields a conserving approximation. In addition to the Hartree and Fock contributions, we consider the electronic self-energy  
\beq{
	\Sigma_{\vk,\alpha\beta}^{\text{GW}}(t, t')=\I \frac{1}{N}\sum_\vq G_{\vk-\vq,\alpha\beta}(t,t') W_{\vq,\alpha\beta}(t,t'),
}
where $W_{\vq,\alpha\beta}$ is the screened interaction, whose self-energy is given by the polarization 
\beq{
	\Pi_{\vq, \alpha\beta}(t, t')=-\I \frac{1}{N}\sum_\vk G_{\vk+\vq, \alpha\beta}(t,t') G_{\vk, \beta\alpha}(t',t).
}
We obtain a closed set of equations by considering the Dyson equation relating the bare and screened interactions, 
\beq{
\label{eq:Wq}
	\hat W_{\vq}=\hat V_{\vq}+\hat V_{\vq}*\hat \Pi_{\vq}*\hat W_{\vq},
}
where $*$ marks the convolution on the Kadanoff-Baym contour and the hat symbol indicates $2\times 2$ matrices in orbital space. Equation~(\ref{eq:Wq}) is valid in the density-density approximation. General interactions would require a two-particle basis.\cite{nilsson2017}

The interaction vertex $\hat V_\vq$, with $V_{\vq,\alpha\beta}=V_\vq$, is instantaneous in time, i.e. $\hat V_\vq(t,t')=\hat V_\vq \delta_{\cal{C}}(t,t')$ with $\delta_{\cal{C}}$ denoting the contour $\delta$-function. Thus, in the  actual implementation, we treat it separately. In practice, we define the charge susceptibility as
\beq{
\hat \chi_\vq=\hat\Pi_\vq + \hat\Pi_\vq *\hat V_{\vq} * \hat\Pi_\vq + \ldots
}
and numerically compute it by solving the integral equation \cite{schueler2020}
\beq{
	\hat\chi_\vq - \hat\Pi_\vq *\hat V_\vq * \hat\chi_\vq =\hat\Pi_\vq.
}
The effective interaction is then determined as $\hat W_\vq=\hat V_\vq+ \hat V_\vq*\hat\chi_\vq*\hat V_\vq.$

In order to calculate the spectral functions, we use traces over the orbital indices, i.e.
\bsplit{
	G_\mathrm{loc}&=\frac{1}{N}\sum_\vk G_\vk \quad \text{with} \quad G_\vk=\Tr \hat G_\vk . 
}

The local spectral functions at time $t_p$ are calculated from the forward-Fourier transformation with respect to $t'$ over a time interval of length $t_\text{max}$,
\bsplit{ 
\label{eq:Aloc}
	A_\text{loc}(t_p,\omega)&=-\frac{1}{\pi}\text{Im}\int_{t_p}^{t_p+t_\text{max}} dt' G_\mathrm{loc}^{R}(t_p,t') e^{-i\omega (t'-t_p)}.
}
For the $\bf{k}$-dependent spectral functions, we apply instead a backward-Fourier transformation 
with respect to $t$:

\bsplit{ 
\label{eq:Ak}
	A_\vk(t_p,\omega)=\frac{1}{\pi}\text{Im}\int_{t_p}^{t_p+t_\text{max}} dt G_\vk^{R}(t,t_p+t_\text{max}) e^{i\omega (t_p+t_\text{max}-t)}.
}

\begin{figure}[hb]
\centering
\includegraphics[width=0.9\columnwidth]{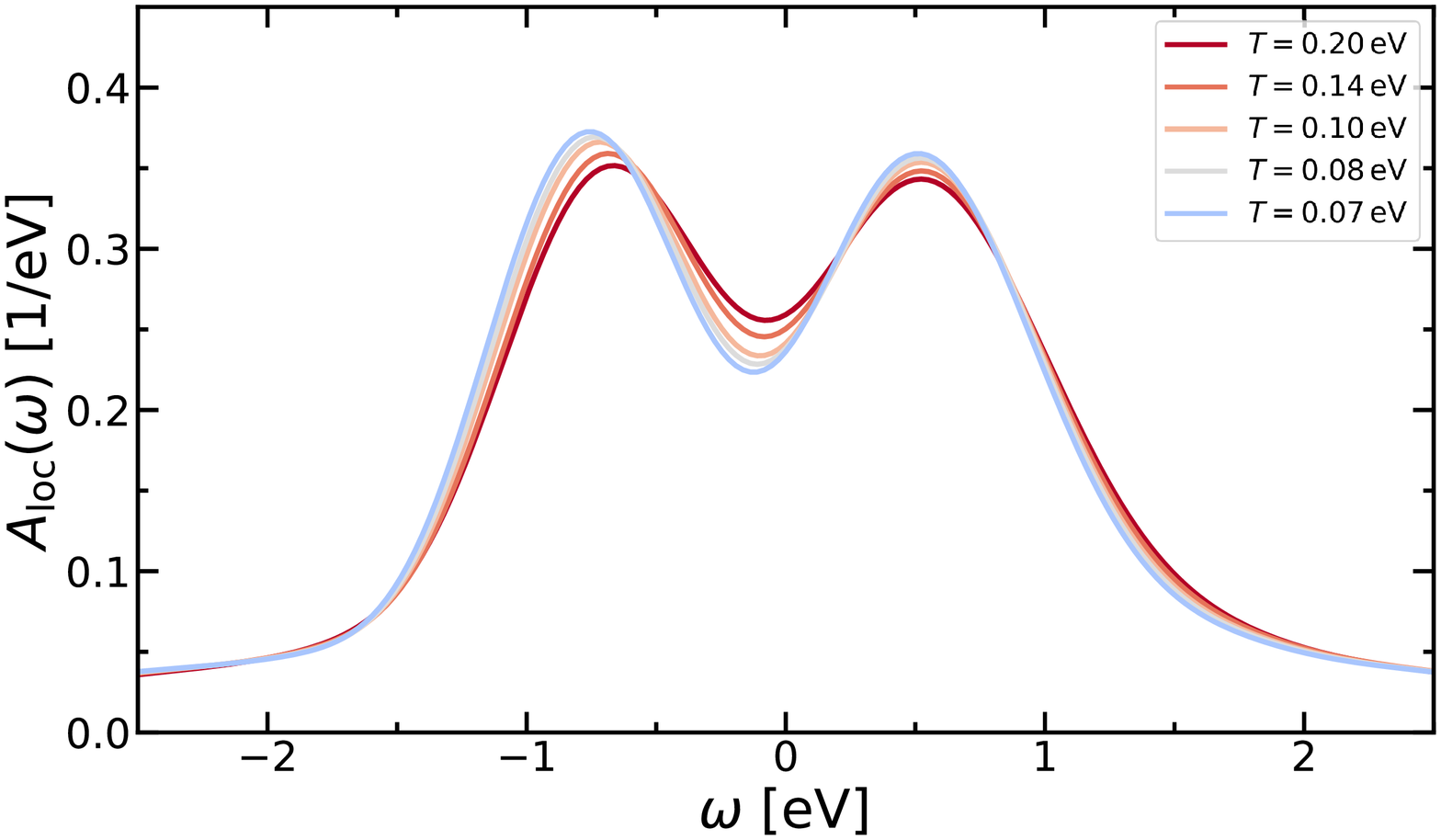}
\caption{Local equilibrium spectral function calculated from the Fourier transformation of $G_\text{loc}^R(t,t')$ with $t_\mathrm{max}=6.8\,$fs for different temperatures. The Dirac behavior near $\omega=0\,$eV is smeared out by the relatively short Fourier window and interaction effects. 
}
\label{fig:GlocT}
\end{figure}

\section{Equilibrium properties}
\label{sec:EqResults}

\subsection{Bandstructure}

Figure~\ref{fig:BandStruct}(a) shows the correlated electronic structure of BaNiS$_2$ (Eqs.~\eqref{eq:H}-\eqref{eq:parm}) along the momentum path $\Gamma=$$(0,0)\to$ M$=$$(\pi,\pi)\to$ X$=$$(\pi,0)\to$ $\Gamma=$$(0,0)$, while the spectral function at the Fermi energy, $A_{\bf k}(\omega=0)$, is shown as a function of ${\bf k}=(k_x,k_y)$ in Fig.~\ref{fig:BandStruct}(b). The band dispersion is calculated from $A_{\bf k}(\omega)$ with $t_\mathrm{max}=6.8\,$fs. This relatively short Fourier window leads to a broadening of the bands. However, we choose here the same value as in the later nonequilibrium simulations with pump pulse, where the time window is restricted by the limited maximum simulation time. 
As one can see, a Dirac cone appears along 
the {\bf k}-path from $(0,0)$ to $(\pi,\pi)$. The Fermi surface consists of four very small Fermi pockets encircling the Dirac points, which are clearly visible in Fig.~\ref{fig:BandStruct}(b) in the form of four spots with strong intensity, in agreement with Refs.~\onlinecite{nilforoushan2019a, nilforoushan2019b}. A consequence of the Dirac dispersions is an approximately linear increase in the density of states away from the Fermi energy, as indicated by the (strongly broadened) local spectral functions plotted in Fig.~\ref{fig:GlocT}.

In order to illustrate renormalization effects on the band dispersion due to different self-energy contributions, we plot in Fig.~\ref{fig:BandStruct}(a) the band structure obtained by keeping only the Hartree-Fock (HF) terms of the full $GW$ self-energy (red dashed lines) and the ``quasi-particle GW" (QPGW)
peak position (red solid lines). The
HF contribution leads to a widening of the band structure compared to its non-interacting counterpart (green dashed line). 
On the other hand, including additionally the $\Sigma^\mathrm{GW}_{\vk}$ self-energy contribution results in a flattening of the band dispersion. The QPGW bands in Fig.~\ref{fig:BandStruct}(a) corresponds to the  
poles of the quasi-particle $GW$ spectral function~\cite{damascelli2003}
\begin{equation}
\label{Eq:PeakPos}
	A_{\vk}(\omega) \propto \frac{1}{\omega-\epsilon_{\vk}+\mu-\Sigma^\mathrm{H}-\Sigma^\mathrm{F}_{\vk}-\mathrm{Re}\Sigma^\mathrm{GW}_{\vk}(E_{\vk})},
\end{equation}
where $E_\vk$ is the self-consistent solution of the equation 
\beq{
	E_{\vk} = \epsilon_{\vk}-\mu+\Sigma^\mathrm{H}+\Sigma^\mathrm{F}_{\vk}+\mathrm{Re}\Sigma^\mathrm{GW}_{\vk}(E_{\vk}).
	\label{Eq:PeakPos1}
}

\subsection{Temperature dependence}

An interesting question concerns the temperature dependence of the screening in this Dirac semi-metal, since this has an important effect on the correlated electronic structure. We analyze here the fully screened interaction $\hat W_\vq$ (Eq.~\eqref{eq:Wq}) computed for different temperatures $T$, using the cutoff $k_{z,\text{cut}}I_c=0.23$. 
This function 
is related to the frequency-dependent dielectric constant by $\hat W_\vq = \hat V_\vq * \hat\epsilon_\vq^{-1}$ (with $\hat \epsilon_\vq = 1-\hat V_\vq * \hat\Pi_\vq $).

\begin{figure}[ht]
\centering
\includegraphics[width=0.93\columnwidth]{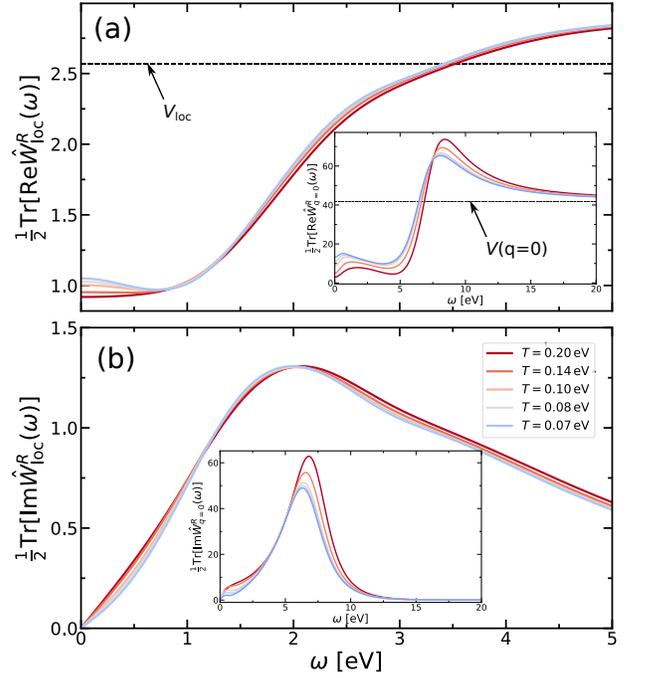}
\caption{Screened interaction $W$ calculated for different temperatures. Panels (a) and (b) show the real and imaginary part of the local component $W_\text{loc}(\omega)$, respectively. The inset shows the corresponding 
$q=0$ components.
}
\label{fig:WLocTdep}
\end{figure}

In Fig.~\ref{fig:WLocTdep}(a), we plot the real part of the screened interaction $\frac{1}{2}\Tr[\mathrm{Re} \hat W_\text{loc}]$
as a function of $T$. For comparison, the local component of the `bare' interaction 
$V_\text{loc}=\frac{1}{2}\Tr[\frac{1}{N}\sum_\vq\hat V_\vq]$
 is shown by the dashed line. As one can see,  
the local 
component of the static screened interaction increases with decreasing temperature, i.e. the screening becomes less effective. The $q\equiv|{\bf q}|=0$ component (inset of Fig.~\ref{fig:WLocTdep}(a)) suggests that this comes primarily from the temperature dependence of the strongly screened long-range interaction, since the static value of the long-range interaction increases significantly as temperature is reduced. This is a characteristic feature of Dirac semi-metals such as BaNiS$_2$, whereas in ``conventional" metals with a large density of states at the Fermi level, one finds the opposite temperature-dependent screening behavior (see Appendix~\ref{sec:EqMetal}).

\begin{figure}[b]
\centering
\includegraphics[width=1.0\columnwidth]{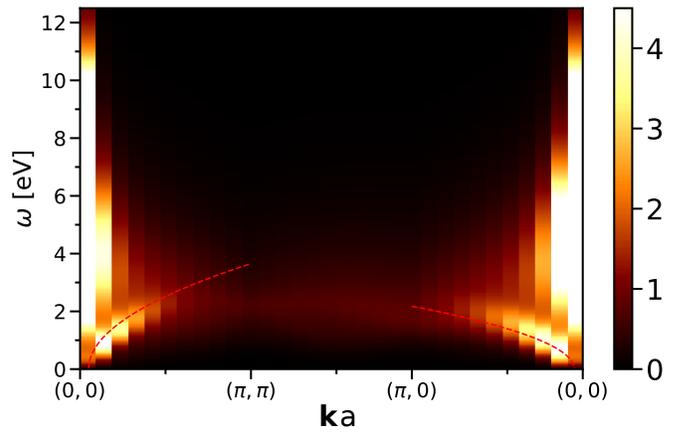}
\caption{Momentum-resolved imaginary part of the screened interaction 
$\frac{1}{2}\Tr[\text{Im}\hat{W}_\vk(\omega)]$ computed at $T=0.07$ eV. 
The red dashed line shows the dispersion of the long wavelength plasmon. Here, we use the formula for the $2d$ conical model with Coulomb interactions from Ref.~\onlinecite{hwang2007}, and the parameters appropriate for BaNiS$_2$. 
}
\label{fig:ImWwk}
\end{figure}

Second, in Fig.~\ref{fig:WLocTdep}(b), we plot the imaginary part $\frac{1}{2}\Tr[\text{Im}\hat W_\text{loc}]$ of the screened interaction, which is related to the real part by the Kramers-Kronig formula. Here, we see a  pronounced peak around $\omega\approx 1.5\,$eV, which corresponds to the valence-conduction band splitting away from the Dirac cones,  
as may be deduced from Fig.~\ref{fig:GlocT}. 
This peak shifts to slightly lower energies as $T$ is decreased, consistent with the larger $\text{Re}[W_\text{loc}]$ at $\omega\gtrsim 2$.   
Similarly, we observe a small peak shift in the long-ranged screened interaction (see the inset of Fig.~\ref{fig:WLocTdep}(b)) at $\omega\approx 6\,$eV. This peak can be associated with single-particle excitations between side bands of $A_{\bf k}(\omega)$ near ${\bf k} = 0$. 
For $\omega\lesssim 1$, in the energy region dominated by the Dirac cones, we see a clear reduction of $\frac{1}{2}\Tr[\text{Im}\hat W_\text{loc}]$ with decreasing $T$, which implies a smaller absolute value of the polarization (see Appendix \ref{sec:PolFct}). This temperature dependence of the polarization is a hallmark of Dirac systems 
and results in the unusual temperature evolution of the screened interaction seen in panel (a).

Another characteristic property of $2d$ Dirac systems is the appearence of a long wavelength plasmon, which has a smaller plasma frequency in the ${\bf q}\to0$ limit than a conventional $2d$ metal.~\cite{hwang2007} 
This plasmon can be identified in the momentum-resolved spectra $\frac{1}{2}\Tr[\mathrm{Im}\hat W_\vq(\omega)]$ as the small peak at low frequencies (see Fig.~\ref{fig:ImWwk}). For illustration, we additionally indicate by the red dashed line the energy dispersion of the plasmon obtained from the analytical formula in Ref.~\onlinecite{hwang2007}, valid for the conical model with Coulomb interactions, and the parameters for BaNiS$_2$. The Dirac prediction nicely matches our numerical results in the ${\bf q}\to0$ limit. 

In addition to the plasmon, we notice the broad feature centered around $\omega\approx 6\,$eV, which
is associated with the peak shown in the inset of Fig.~\ref{fig:WLocTdep}(b).

\begin{figure}[b]
\centering
		\includegraphics[width=0.95\columnwidth]{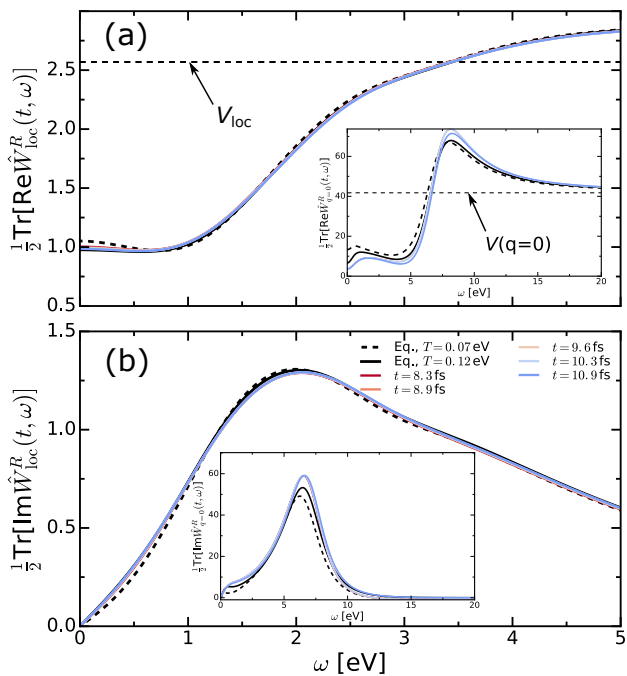}
\caption{Time-dependent (a) real and (b) imaginary part of the local screened interaction $W(\omega)$ after a photo-excitation. 
The equilibrium results at the initial $T=0.07$ and the temperature of the thermalized system ($T=T_\text{eff}=0.12$) are shown by dashed and solid black lines, respectively. In the main panels, the blue lines are on top of the solid black lines. 
}
\label{fig:NoneqWLocTdep}
\end{figure} 

\section{Non-Equilibrium results}
\label{sec:NonEqResults}

\subsection{Pulse excitation}

In this section, we investigate the time-dependent modifications in the electronic structure and screening properties of BaNiS$_2$ after a photoexcitation. We simulate the photo-excitation of the system with a short light pulse by direct dipole allowed transitions between the orbitals contributing to the Dirac cone (see Eq.~\eqref{eq:Hdip}). For the pump pulse we choose the parameters $\omega=1.5\,$eV, $A_0=0.15$, $\tau\approx 2.3\,$fs, and $t_0=3.6\pi/\omega$
, unless otherwise specified. While this pulse frequency is consistent with the experiments in Ref.~\onlinecite{nilforoushan2019a}, the pulse duration is shorter (it contains only about five cycles), due to computational limitations in the accessible time range. Before the pump, the system is in equilibrium at $T=0.07\,$eV, where its low-energy properties are dominated by the Dirac physics.  
The following results are obtained using a  $20\times20$ grid in momentum space. A finer ${\bf k}$ grid does not significantly change the results. 

 \begin{figure}[b!]
\centering
\includegraphics[width=0.99\columnwidth]{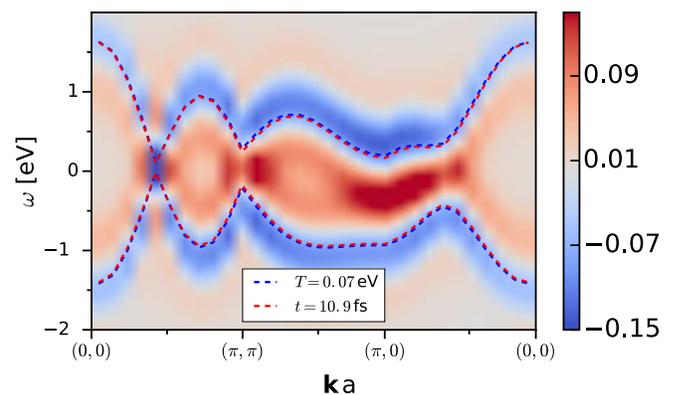}
\caption{Difference $\Delta A_k$ between the spectral function of the photo-excitated system ($t\approx 10.9\,$fs) and the initial equilibrium spectral function ($T=0.07\,$eV). The dashed lines indicate the QPGW peak positions from Eq.~\eqref{Eq:PeakPos} for the initial equilibrium state (blue dashed line) and after the photo-excitation (red dashed line). 
Both spectral functions are calculated with $t_\text{max}=6.8\,$fs.
}
\label{fig:NoneqImGlesVeffTdep}
\end{figure} 

\begin{figure*}[ht]
\centering
	\begin{tabular}{c}
\includegraphics[width=1.\textwidth]{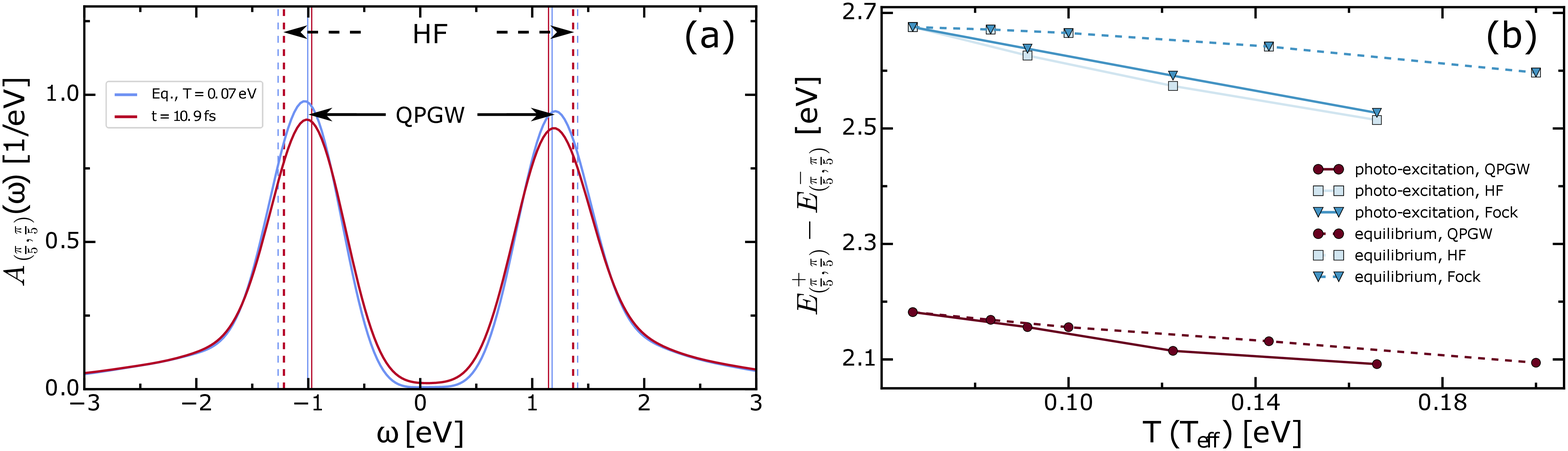}
	\end{tabular}
\caption{(a) Time-dependent ${\bf k}$-resolved spectral function calculated at $\vk a=(\pi/5,\pi/5)$. The blue line corresponds to the equilibrium spectrum at $T=0.07\,$eV and the red line to the nonequilibrium spectrum after the pulse at $t\approx 10.9\,$fs. Solid (dashed) vertical lines indicate the positions of the QPGW bands (bands which take into account the Hartree-Fock contribution to the full GW self-energy). 
(b) Difference between the QPGW energies of the upper (``+") and lower (``-") band at ${\bf k}a=(\pi/5,\pi/5)$ in equilibrium (dark red dashed line with dots) and after a photo-excitation (dark red solid line with dots) as a function of $T$ or $T_\text{eff}$. In addition, we plot the analogous results taking into account only the Fock (blue lines with triangles) and Hartree-Fock (light blue lines with squares) contributions to the full $GW$ self-energy. The effective temperature after the photo-excitation is estimated from the total energy. 
}

\label{fig:NoneqLocImGretkVeffTdepE02Kzcut}
\end{figure*} 

\subsection{Nonequilibrium screening}
\label{subsec_screening}

First, we analyze the screening effects after the photo-excitation. In Fig.~\ref{fig:NoneqWLocTdep}, we plot the real and imaginary parts of the time-dependent screened interaction. As one can see, the static screened interaction is slightly 
reduced after the photo-excitation (Fig.~\ref{fig:NoneqWLocTdep}(a)), while the peak in the  imaginary part is slightly shifted from $\omega\approx 1.8\,$eV to higher energies. This is qualitatively similar to the effect of heating in equilibrium (c.f. Fig.~\ref{fig:WLocTdep}). For comparison, we plot in the same figure also the equilibrium result for $T = 0.12\,$eV, 
which is the temperature $T_\text{eff}$ of the thermalized system (black solid line). The low-frequency effective interaction $\frac{1}{2}\Tr[\text{Re}\hat W^R_\text{loc}(t,\omega=0)]$  
after the pulse resembles the equilibrium result at $T=0.12\,$eV. On the other hand, the peak in $\frac{1}{2}\Tr[\text{Im}\hat W^R_\text{loc}]$ is shifted to slightly higher frequencies than in equilibrium. This indicates some differences in the broadening or energy position of the bands, but these differences are rather small. 
The inset shows that the thermalization of the $q=0$ component is slower than in the case of the local component. 
 
An analysis of the ${\bf q}$ dependent nonequilibrium results shows that the ${\bf q}\rightarrow0$ plasmon and the broad peak near $\omega\approx 6\,$eV are enhanced, while for larger $q$ the pulse induces mainly a broadening of the features, again in qualitative agreement with the effect of increasing temperature in equilibrium. 
Hence, from the nonequilibrium behavior of $W$ it appears that the photo-induced changes in the screening environment are essentially equivalent to the changes induced by heating. 
If significant differences in the band renormalization between photo-doped and heated systems are found, 
it is hence unlikely that the dominant effect is the nonthermal screening environment. 

\subsection{Nonequilibrium spectral function}

To study the time- and momentum-resolved spectral function, we calculate $A_{\bf k}(t,\omega)$ in analogy to Eq.~\eqref{eq:Aloc}. In Fig.~\ref{fig:NoneqImGlesVeffTdep}, we plot the difference between the photo-excited ($t\approx 10.9\,$fs) and the  equilibrium spectrum ($T=0.07\,$eV). Even though the spectral resolution is limited because of the short Fourier window ($t_\mathrm{max}=6.8\,$fs), it is obvious that the spectral weight in the momentum region associated with the Dirac cone is shifted towards $\omega = 0$, which implies a flattening of the Dirac dispersion. In addition, the figure shows a significant broadening of the bands in the photo-excited state, consistent with the results from a recent $GW$+DMFT study.\cite{Golez2019}

To clearly illustrate the flattening of the Dirac cones, we plot in Fig.~\ref{fig:NoneqLocImGretkVeffTdepE02Kzcut} the spectra of the equilibrium and photo-excited system for $\vk a=(\pi/5,\pi/5)$. While the photo-excitation induces both shifts and broadenings of the spectra, the peak positions are slightly shifted towards $\omega =0 $. 
This result, which is not sensitive to the cut-off value $k_{z,\text{cut}}$,\footnote{The only effect of $k_{z,\text{cut}}$ is a rescaling of the dispersions.} is consistent with a flattening of the Dirac cone. 

For a better visualisation of the energy shifts, we indicate in the same figure the QPGW peak positions $E_\vk$ from Eq.~\eqref{Eq:PeakPos} by vertical solid lines.
We furthermore show by vertical dashed lines the peak positions obtained by keeping only the Hartree-Fock contribution to the full $GW$ self-energy. The photo-induced change of the interorbital Hartree-Fock contribution along the $\Gamma\to M$ direction can be calculated as (see Appendix~\ref{sec:SC1}):
\beq{
\label{eq:dShf}
	\Delta\Sigma^F_{\vk,\alpha\alpha} + \Delta\Sigma^H_{\alpha\alpha} = -\frac{1}{N}\sum_\vq V(\vq)\Delta \rho_{\vk-\vq,\alpha\alpha} (t),
}
where $\Delta\rho$ represents the photo-induced change in the charge density distribution. Since the photo-excitation reshuffles charge from orbital $\alpha=1$ to orbital $\alpha=2$, corresponding to $\Delta\rho_{\alpha\alpha}<0$ ($>0$) for the states below (above) FL, 
it results in the flattening of the effective bandstructure obtained by keeping the Hartree-Fock terms of the $GW$ self-energy.   
This is clearly visible by looking at the dashed lines in Fig.~\ref{fig:NoneqLocImGretkVeffTdepE02Kzcut}(a). 
Moreover, we can see that the bandstructure which accounts only for the Hartree-Fock contributions responds more strongly 
to the photo-doping than the full QPGW bandstructure, which indicates that the non-retarded (Hartree-Fock) and retarded self-energy diagrams have opposite effects on the correlated electronic structure, and that the former dominate the band shifts. 

Similar effects as in the photo-doped system can be found by increasing the temperature in equilibrium, as is shown in Appendix~\ref{sec:BS}.
However, 
the nonthermal population created by the photo-doping leads to a stronger flattening than a simple heating. 
For a quantitative analysis of the corresponding band renormalizations, we plot in Fig.~\ref{fig:NoneqLocImGretkVeffTdepE02Kzcut}(b) 
the energy difference between the upper and lower band at ${\bf k}a=(\pi/5,\pi/5)$
at different temperatures and after the photo-excitation. 
For a proper comparison between equilibrium and nonequilibrium results, we estimate the effective temperature $T_\text{eff}$ after pumping from the change in the total energy, i.e. we search for the equilibrium system with $T=T_\text{eff}$ and the same total energy as the photo-excited system. 
In Fig.~\ref{fig:NoneqLocImGretkVeffTdepE02Kzcut}(b), we show that the flattening of the cone 
is qualitatively similar in the photo-doped and heated system, but the effect of photo-doping is significantly stronger. 
Comparing separately the effects of the Fock, Hartree-Fock, and the full $GW$ self-energy on the band shifts, we see that the nonthermal contribution to the Dirac cone flattening is primarily due to the Fock term. The enhanced band-flattening in the photo-doped state and the importance of the Fock contribution represent the main results of this study. Together with the analysis in Sec.~\ref{subsec_screening} we may conclude that rather than an effect of nonthermal screening, the pronounced changes measured by tr-ARPES in photo-excited BaNiS$_2$ originate from the Fock exchange.

\section{Summary}
\label{sec:Summary}

We presented a theoretical study of the effect of photo-excitation on the electronic structure of BaNiS$_2$ and the possibility of a photo-induced Dirac cone flattening in this quasi two-dimensional Dirac material. Our study combined an {\it ab-initio} inspired model for BaNiS$_2$ with a nonequilibrium $GW$ treatment of the photo-induced dynamics. This advanced methodology allowed us to study both the screening properties and interactions in the nonequilibrium state and their effect on the nonequilibrium electronic structure. 

Calculating the effective screened interaction at different temperatures in equilibrium, we demonstrated a non-trivial screening enhancement with increasing temperature, which is a hallmark of Dirac systems. In particular, we showed that this effect comes primarily from a strong screening of the long-range part of the interaction, since the $q=0$ component of the screened interaction decreases strongly with increasing temperature. This can be traced back to the properties of the polarization function, which exhibits a positive temperature slope within the Dirac region, in contrast to a conventional metal.  

Our nonequilibrium simulations of BaNiS$_2$ revealed both an effective heating of the solid after a photo-excitation, and photo-induced nonthermal effects. By computing the effective temperatures of the photo-doped systems, and from comparisons with thermal data, we concluded that the changes in the screening environment can 
to a large extent be
explained by the heating effect.
On the other hand, the band renormalization effects are much stronger in the photo-doped state than in thermalized systems at the corresponding effective temperature.

A separate analysis of the Hartree and Fock contributions to the QPGW bandstructure showed 
that this nonthermal effect is mainly driven by the Fock exchange and the out-of-equilibrium charge distribution. Our study demonstrates that a nontrivial combination of heating, modifications in the screening environment, as well as band shifts due to the Fock term are all relevant to describe photo-doped Dirac semi-metals. This exemplifies the usefulness of unbiased nonequilibrium many-body simulations for the interpretation of ultrafast time-resolved experiments on correlated solids. 

\begin{acknowledgments}
This work was supported by ERC Consolidator Grant No. 724103 (NB, PW), and Swiss National Science Foundation Grant No. 200021\_196966 (PW). The calculations have been performed on the Beo04 cluster at the University of Fribourg. DG is supported by Slovenian Research Agency (ARRS) under Program J1-2455 and P1-0044. 
\end{acknowledgments}

\appendix

\begin{figure}[b]
\centering
		\includegraphics[width=0.75\columnwidth]{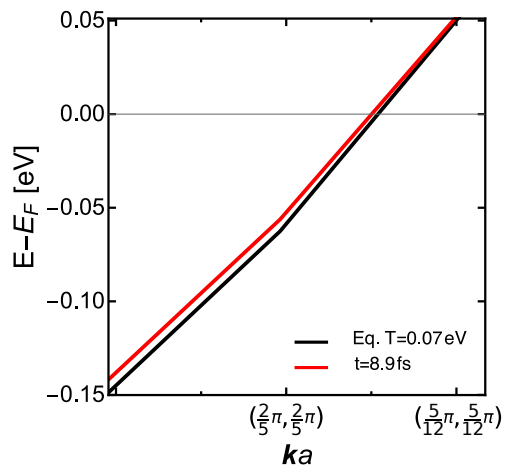}

\caption{Dirac cone flattening of the left branch obtained from the $GW$ real-time simulations. The black curve shows the initial equilibrium dispersion and the red curve the dispersion after the photo-doping pulse. 
These simulations are performed with the dielectric constant $\kappa_0=10$. The initial temperature is $T=0.07\,$eV.
}
\label{fig:Theo}
\end{figure}

\section{Flattening of the Dirac cone for $\kappa_0=10$}
\label{sec:kappa10}

To illustrate the flattening of the Dirac cone in BaNiS$_2$ after a photo-excitation in a model with a more realistic screening parameter, we performed a nonequilibrium $GW$ simulation with $\kappa_0=10$ (see Eq.~\eqref{eq:Vqkz}). The initial temperature is set to $T=0.07\,$eV and the system is excited by a pump pulse with parameters $\omega=1.5\,$eV, $A_0=0.1$, $\tau=1.3\,$fs. In Fig.~\ref{fig:Theo} we show the results of the simulations, where the black line illustrates the left branch of the Dirac cone in equilibrium and the red line shows the dispersion after a photo-excitation (measured at $t=8.9\,$fs). Even though the limited {\bf k}-point resolution makes a direct comparison difficult, one can see by comparison to Fig.~\ref{fig:Exp} that the photo-induced Dirac cone flattening in our simulation is comparable to the flattening observed in the tr-ARPES experiments.

\section{Temperature dependence of the band structure}
\label{sec:BS}

\begin{figure}[t!]
\centering
		\includegraphics[width=0.95\columnwidth]{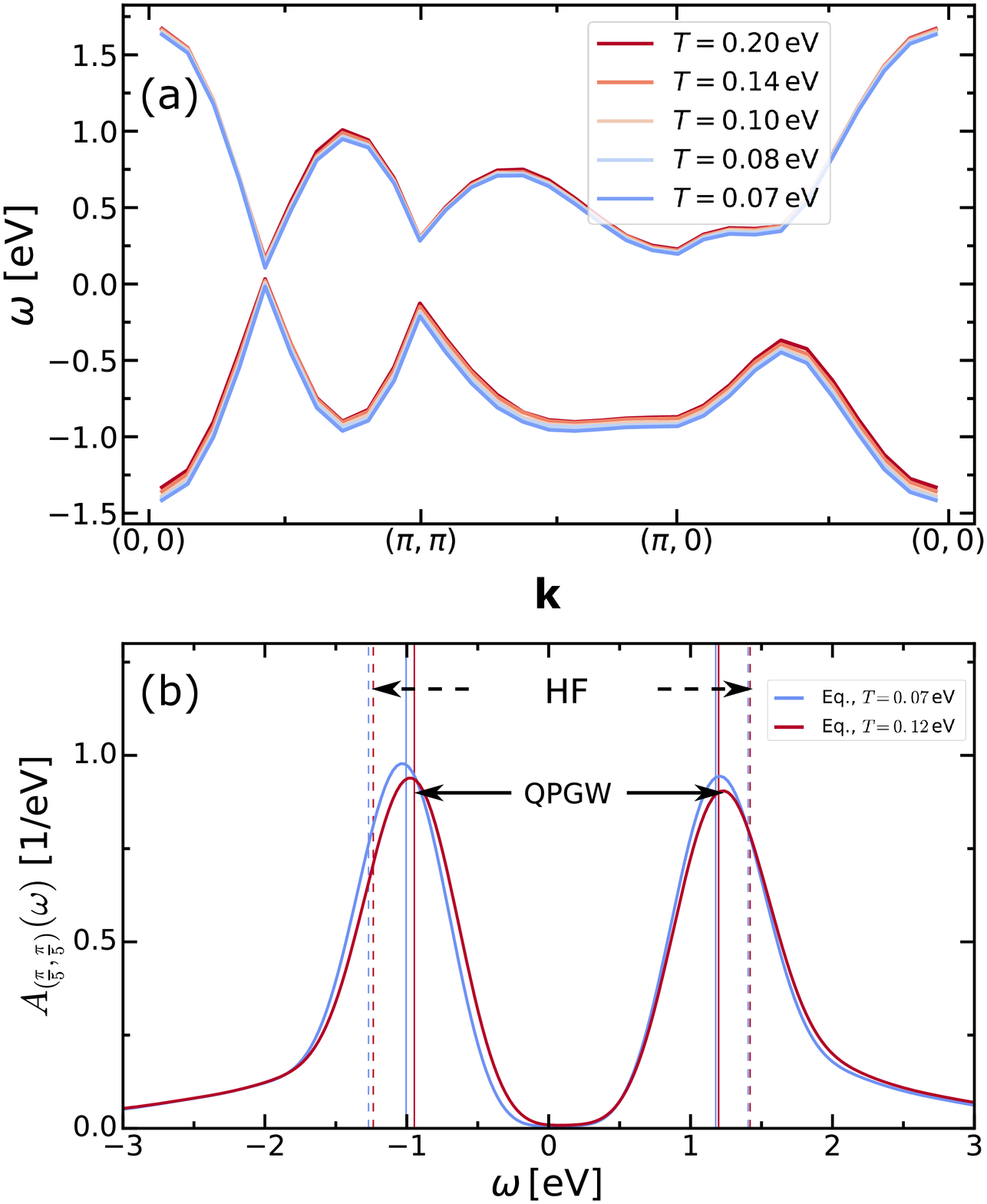}
\caption{(a) Temperature dependence of the QPGW band structure calculated from Eq.~\eqref{Eq:PeakPos}. (b) ${\bf k}$-resolved spectral function calculated at $\vk a=(\pi/5, \pi/5)$ for $T=0.07\,$eV (blue line) and $T=0.12\,$eV (red line).
}
\label{fig:EqQPGW}
\end{figure}

In Fig.~\ref{fig:EqQPGW}(a) we show the QPGW peak positions at different temperatures (see Eq.~\eqref{Eq:PeakPos}). As one can see, the band structure of BaNiS$_2$ does not significantly change as temperature is increased. However, at certain ${\bf k}$ points, 
one finds a small broadening of the upper band at higher temperatures, whereas the lower band shows a simultaneous shift towards $\omega=0$. 
To illustrate this, we plot in Fig.~\ref{fig:EqQPGW}(b) $A_{\bf k}(\omega)$ for $\vk a=(\pi/5,\pi/5)$ at $T=0.07\,$eV and 0.12$\,$eV ($T_\mathrm{eff}$ from Sec.~\ref{sec:NonEqResults}). As one can see, increasing the temperature leads to a broadening of the momentum-resolved spectral functions. However, the peak position of the upper band does not change with temperature, whereas the lower band shifts toward $\omega=0$.

\section{Photo-induced changes in Hartree-Fock}
\label{sec:SC1}

The photo-doping of the system leads to the time-dependent changes in the local (loc) charge density distribution with respect to the equilibrium (eq) situation. For simplicity, we focus only on the diagonal contributions:
\beq{
\label{eq:dSht}
\rho_{\text{loc}, \alpha\alpha}(t) = \rho_{\text{loc}, \alpha\alpha}(\text{eq}) + \Delta \rho_{\text{loc},\alpha\alpha}(t).
}
The changes of the charge density distribution during the pulse are
\begin{equation}
	\Delta\rho_{\text{loc}\alpha\alpha} = \left\{
		\begin{array}{c c}
			>0 & \text{for } \alpha=2,\\
			<0 & \text{for } \alpha=1.\\
		\end{array}
	\right.
\end{equation}
From this, we can obtain the changes in the Hartree self-energy due to the photo-excitation. 
Using the definition of the Hartree self-energy: 

\begin{figure}[b!]
\centering
\includegraphics[width=1.0\columnwidth]{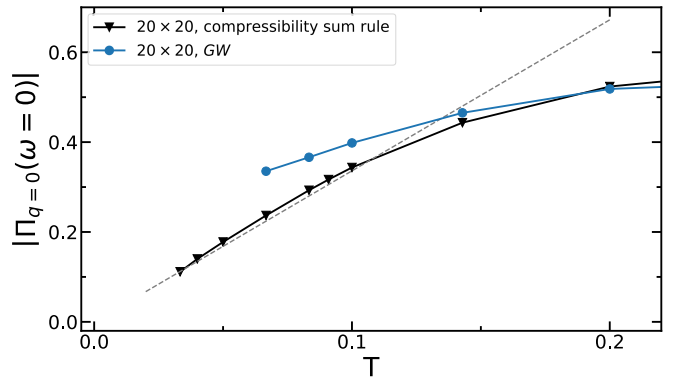}
\caption{$q=0$ component of the polarization function in the static limit, $\Pi_{q=0}(\omega=0)$, calculated at different temperatures. Blue dots are obtained from equilibrium $GW$ real-time calculations, whereas the black triangles show the result from the compressibility sum rule. The gray dashed line shows the result of the analytical formula derived in Ref.~\onlinecite{nilforoushan2019a} in low temperature limit. 
}
\label{fig:PiStat}
\end{figure}

\begin{figure*}[htb!]
\centering
		  \includegraphics[width=0.95\textwidth]{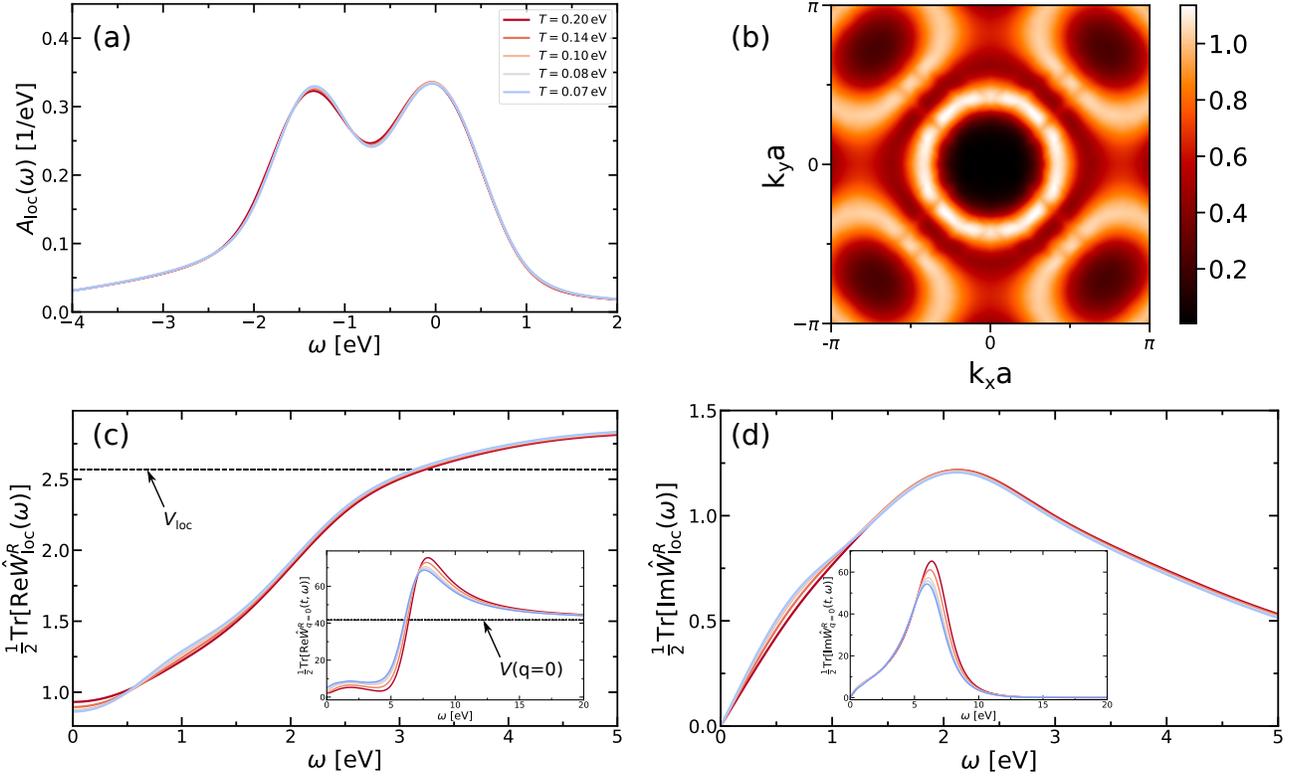}
\caption{(a) Temperature-dependent local spectral of a conventional metal (stronly doped BaNiS$_2$). (b) Corresponding $A_\vk(\omega=0)$ for $T=0.07\,$eV plotted in the first Brillouin zone. (c) Real and (d) imaginary part of $W_\text{loc}(\omega)$ calculated for different temperatures (see labels in panel (a)). The inset shows the corresponding $q=0$ components. 
}
\label{fig:EqAkMetal}
\end{figure*} 

\bsplit{
\Sigma^H_{\text{loc}, \alpha\beta}(t)=&\left(\delta_{\alpha\beta} V(q=0) \rho_{\text{loc},\bar\alpha\bar\alpha}(t)\right.\\
&\left.+ \delta_{\alpha\beta} [V(q=0)-V_\text{loc}] \rho_{\text{loc}, \alpha \alpha}(t)\right)\\
=& \delta_{\alpha\beta} [V(q=0)\rho_\text{tot} -V_\text{loc}\rho_{\text{loc}, \alpha \alpha}(t)]
}
with $\rho_\text{tot}=\rho_{\text{loc}, 00} + \rho_{\text{loc}, 11}$, we get
\beq{
\Delta\Sigma^H_{\alpha\alpha}(t)\equiv \Sigma^H_{\alpha\alpha}(t) - \Sigma^H_{\alpha\alpha}(\text{eq}) = - V_\text{loc} \Delta \rho_{\text{loc},\alpha\alpha}(t).
}
In other words, the Hartree component leads to a narrowing of the band width after the photo-excitation. Additionally, we note that for a fixed value of the long-range Coulomb interaction $V(q=0)$ the local interaction $V_\text{loc}=1/N\sum_{\bf q} V({\bf q})$ is larger for smaller $\kappa_0$. Hence, the effect of a narrowing band width due to the photo-induced Hartree shift should be larger for smaller $\kappa_0$.

Now, let us focus on the photo-induced changes in the interband Fock term. Using Eq.~\eqref{eq:Fock} we get
\bsplit{
\label{eq:dSf}
\Delta\Sigma^F_{\vk,\alpha\alpha} &= \Sigma^F_{\vk,\alpha\alpha}(t) - \Sigma^F_{\vk,\alpha\alpha}(\text{eq}) \\
&= -\frac{1}{N}\sum_\vq V(\vq)\Delta \rho_{\vk-\vq,\alpha\alpha} (t) + V_\text{loc} \Delta \rho_{\text{loc},\alpha\alpha}(t).
}
Thus, from Eq.~\eqref{eq:dSht} and Eq.~\eqref{eq:dSf} we obtain
\beq{
	\Delta\Sigma^F_{\vk,\alpha\alpha} + \Delta\Sigma^H_{\alpha\alpha} = -\frac{1}{N}\sum_\vq V(\vq)\Delta \rho_{\vk-\vq,\alpha\alpha} (t), 
}
where $\Delta\rho_{\vk,\alpha\alpha}<0$ ($>0$) for the states below (above) the FL. 
It follows that along the relevant direction $\Gamma\to M$ in momentum space, 
the photo-induced changes in the Hartree-Fock contributions to the $GW$ self-energy lead to a flattening of the bands directly after the pulse. This effect should be larger for interactions with smaller $\kappa_0$.

\section{Polarisation function}
\label{sec:PolFct}

Here, we calculate the temperature dependence of the local component of the static polarization function, which is plotted in Fig.~\ref{fig:PiStat}. As one can see, $\Pi_{q=0}(\omega=0,\beta)$ computed from the compressibility sum rule
\beq{
	\Pi_{q=0}(\omega=0, \beta)=\int_{-\infty}^{\infty} d\epsilon \left(\frac{\partial f(\epsilon)}{\partial \epsilon}\right) D(\epsilon),
}
 shows a linear increase below $T\approx0.1$, which can be described by the analytical expression for Dirac cone dispersions derived in Ref.~\onlinecite{nilforoushan2019a} (gray dashed line). In the above equation, $f$ denotes the Fermi function and $D(\epsilon)$ the density of states. 
 
 We note that for the comparison with our results we choose a definition of the polarization function without a negative sign. At higher temperatures one observes deviations from this linear behavior, which originate from thermal excitations beyond the Dirac region. 
Within our $GW$ approximation, the Dirac behavior of the polarization ($\Pi_{q=0}(\omega=0) = \Tr[\mathrm{Re}\hat\Pi_{q=0}(\omega=0)]$) is not entirely reproduced (see blue line in Fig.~\ref{fig:PiStat}). The reason is that $GW$ breaks the compressibility sum rule, which can be restored by including vertex corrections beyond the $GW$ approximation.~\cite{takada2001} However, at $T\lesssim 0.1$ we still see a qualitatively similar temperature dependence with a positive temperature slope, which is qualitatively different from the temperature dependence for conventional metals.  

\section{Equilibrium screening in a conventional metal}
\label{sec:EqMetal}

To illustrate the different screening behavior in a conventional metal, compared to the Dirac semi-metal BaNiS$_2$, we analyze the temperature dependence of the screened interaction in a highly doped system ($\approx 50\%$). We consider the model described by Eq.~\eqref{eq:H} and shift the chemical potential away from the Dirac points to approximately the energy of the upper peak in $A_\mathrm{loc}$ (see Fig.~\ref{fig:EqAkMetal}(a)). This results in a large, ring-shaped Fermi-surface, as shown in Fig.~\ref{fig:EqAkMetal}(b). 

The calculation of the real part of the screened interaction 
$\frac{1}{2}\Tr[\mathrm{Re}\hat W_\text{loc}]$
 at different temperatures in this conventional metallic case yields a reduced screening with increasing $T$ (see Fig.~\ref{fig:EqAkMetal}(c)), and a long-ranged static component which is almost completely suppressed at $T=0.2$ eV. In the imaginary part of $W_\text{loc}$ one finds a pronounced peak around $\omega\approx 2.0\,$eV (see Fig.~\ref{fig:EqAkMetal}(d)), which approximately corresponds to the band splitting, as shown in panel (a). This peak is slightly shifted to lower energies by decreasing $T$. In the long-range part of the interaction (inset) one observes a similar shift of the peak at $\omega\approx 6\,$eV. 
Interestingly, for $\omega\leq 1$ eV, we see a clear enhancement of $\frac{1}{2}\Tr[\mathrm{Im}\hat W_\text{loc}]$ with decreasing temperature. 

These results illustrate 
that the screened interaction (and polarization) in a conventional metal show the opposite temperature behavior from a Dirac semi-metal, such as BaNiS$_2$ (c.f. Fig.~\ref{fig:WLocTdep}).

\newpage
\bibliography{banis}
\end{document}